\documentclass[aps,pra,twocolumn,groupaddress]{revtex4-1}
\usepackage{graphicx}
\usepackage{amsmath,amssymb}
\usepackage{bm}
\usepackage{tikz}
\usetikzlibrary{positioning}

\newcommand{\etal}{{\it et al.\/}}
\newcommand{\idest}{{\it i.e.\/}}
\newcommand{\dd}{{\rm d}}
\newcommand{\comm}[2]{\left[#1,#2\right]}
\newcommand{\ee}[1]{\relax\ifmmode 10^{#1} \else 10$^{#1}$\fi}
\newcommand{\qvalue}[1]{\left\langle #1 \right\rangle}

\renewcommand{\vr}{\textbf{r}}
\newcommand{\vj}{\textbf{j}}
\newcommand{\vn}{\bm{\nabla}}
\newcommand{\ve}{\textbf{e}}
\newcommand{\vv}{\textbf{v}}
\newcommand{\cS}{{\cal S}}
\newcommand{\cL}{{\cal L}}
\newcommand{\nL}{{L}}
\newcommand{\cH}{{H}}
\newcommand{\dkap}{\dot{\kappa}}
\newcommand{\ddkap}{\ddot{\kappa}}
\newcommand{\tkap}{\tilde{\kappa}}
\newcommand{\dtkap}{\dot{\tkap}}
\newcommand{\ddtkap}{\ddot{\tkap}}

\newcommand{\db}{\dot{b}}

\newcommand{\tx}{\tilde{x}}
\newcommand{\ty}{\tilde{y}}
\newcommand{\tz}{\tilde{z}}
\newcommand{\tr}{\tilde{\rho}}
\newcommand{\haak}[1]{\left[ #1 \right]}
\newcommand{\embr}[1]{\left( #1 \right)}
\renewcommand{\eqref}[1]{Eq.~(\ref{#1})}
\newcommand{\eqsref}[2]{Eqs.~(\ref{#1}-\ref{#2})}
\newcommand{\TtwoB}{T^\mathrm{2B}}
\newcommand{\Ri}[1]{R_{#1}(0)}
\newcommand{\wnul}{\omega_\rho}
\newcommand{\inul}{\omega_i}
\newcommand{\znul}{\omega_z}
\newcommand{\dingus}{\Ri{\rho}^2}
\newcommand{\dattum}{\wnul{}^2}
\newcommand{\dittum}{\inul{}^2}
\newcommand{\dottum}{\znul{}^2}
\newcommand{\doubpart}[2]{\frac{\partial #1}{\partial #2}}
\newcommand{\doubfull}[2]{\frac{\dd #1}{\dd #2}}

\newcommand{\hnj}{\hat{n}_j}
\newcommand{\hfj}{\hat{\phi}_j}
\newcommand{\haj}{\hat{a}_j}
\newcommand{\dhaj}{\dot{\hat{a}}_j}
\newcommand{\hpj}{\hat{\pi}_j}
\newcommand{\hajd}{\hat{a}_j{}^\dagger}
\newcommand{\scref}[1]{Sec.~\ref{sc:#1}}

\begin{document}
\title{Dynamics of a Space-Time Crystal in an Atomic Bose-Einstein Condensate}

\author{L. Liao${}^1$}
\author{J. Smits${}^2$}
\author{P. van der Straten${}^2$}
\email[]{p.vanderstraten@uu.nl}
\author{H.T.C. Stoof${}^1$}

\affiliation{$^1$Institute for Theoretical Physics and Center for Extreme Matter and Emergent Phenomena, Utrecht University, PO Box 80.000, 3508 TA Utrecht,The Netherlands\\
$^2$Debye Institute for Nanomaterials and Center for Extreme Matter and Emergent Phenomena, Utrecht University, PO Box 80.000, 3508 TA Utrecht,The Netherlands}
\date{\today}%

\begin{abstract}
A space-time crystal  has recently been observed in a superfluid Bose gas. Here we construct a variational model that allows us to describe from first principles the coupling between the radial breathing mode and the higher-order axial modes that underlies the observation of the space-time crystal. By comparing with numerical simulations we verify the validity of our variational {\it Ansatz}. From the model we determine the requirements for the  observation of the space-time crystal and the Ising-like nature of the symmetry breaking involved. Also, we find the onset and growth rate of the space-time crystal, which can be compared to experiments.
\end{abstract}

\maketitle

\section{Introduction}

The idea of time crystals was proposed by Wilczek in 2012 \cite{wilczek_tc,Wilczek2012b}, where for a system in its ground state the continuous time-translation symmetry is spontaneously broken. The proposal stimulated vigorous debates about the properties of a time crystal \cite{bruno,nozieres}. Time crystals based on the time-dependent correlation functions are proposed by Watanabe~\etal~\cite{oshikawa_no_tc}, and they showed that spontaneously breaking of the continuous time symmetry into a discrete symmetry in the quantum ground state is impossible. However, the possibility of breaking of a discrete time translation symmetry is not ruled out~\cite{schoen_ph_sp_xtal,Sacha,nayak_floq_tc,sondhi1,sondhi2,else,Syrwid2017}. Yao~\etal~\cite{yao} proposed a model of a one-dimensional discrete time crystal occurring in periodically driven spin systems, which is followed by experimental observations of a discrete time crystal in an interacting spin chain of trapped atomic ions \cite{monroe_exp}, and in disordered \cite{lukin_exp} and ordered  magnetic systems~\cite{Rovny,barrett_exp,Autti2018}. Although the experiments clearly showed the breaking of discrete time symmetry, the experiments do not allow for a full theoretical description, since the underlying physics of those spin systems is far too complex. In contrast, Smits~\etal~\cite{Smits2018} reported the observation of a space-time crystal in a superfluid gas, where excitations can be described nearly from first principles. Recently, time crystals have been reviewed in Ref.~\cite{sacha-review}.

In this paper we describe a variational model, that fully describes the experimental findings of Ref.~\cite{Smits2018}. The model is based on the interaction between the radial breathing mode and  higher-order axial modes, which have been observed before and were dubbed a ``Faraday'' wave~\cite{engels_faraday}. Both excitations are fully determined by their density and phase and this quantum hydrodynamical description allows for a complete characterization of the different orders of the coupling. The results of the model are compared to numerical simulations based on the Gross-Pitaevskii (GP) equation, which accurately describes the dynamics in the superfluid. The agreement between the model and numerical simulations shows that our model captures all the relevant dynamics in the superfluid needed to understand the formation and growth of the space-time crystal.

Since our variational model is constructed from first principles, we can extract properties of the space-time crystal, which can be compared to the theoretical proposals for space-time crystals~ \cite{Huang,lustig,WenWei,Mizuta,Russo}. In our model we identify a prethermal state for the space-time crystal, that is only weakly coupled to the ground state of the system and thus allows for its observation in the experiment. We show that the Hamiltonian describing the system has a hidden $\mathbb{Z}_2$ symmetry, which is broken in the crystalline phase.    

The paper is organized as follows. In \scref{experiment} we shortly describe the experimental sequence that leads to the space-time crystal. In \scref{action} we introduce the variational model and define the higher-order axial modes, which are given in terms of Legendre polynomials. The model allows for the description of the coupling between the radial and axial excitations and the second- and third-order interactions are taken into account. In \scref{simul} we test our variational {\it Ansatz} and find satisfactory agreement between the model and simulations for the atomic line density and the atomic flux. In \scref{diag} we derive the evolution equations for the amplitudes of the higher-order axial modes and determine the frequencies of the modes, which are compared to the frequency of their oscillations in the numerical simulations. In \scref{back} we show how the higher-order axial modes lead to a back action on the breathing mode. In \scref{STX} we present a quantized version of our Hamiltonian and derive the quantum properties of the space-time crystal, that has been observed recently~\cite{Smits2018}. Finally we derive in \scref{higher} the onset of the higher-order axial modes and their gain, and argue that the modes can only be excited in a narrow window of the driving amplitude.

\section{The experiment\label{sc:experiment}}

In the experiment~\cite{Stam}  we cool $^{23}$Na atoms using laser and evaporative cooling techniques to a temperature of 300 nK, which is far below the BEC transition temperature of about 1 $\mu$K. In this way the cloud is nearly fully condensed, although there always remains a small fraction of thermal atoms. The atoms are trapped in a cylindrical symmetric harmonic trap with the potential $V(\rho,z)$ given by
\begin{equation}
V(\rho,z)=\frac{1}{2}m\omega_\rho{}^2 \embr{\rho^2+\lambda^2 z^2}, \label{eq1}
\end{equation}
with $m$ the mass of the atoms, $\omega_\rho$ the trap frequency in the radial direction $\rho$ and $\omega_z=\lambda \omega_\rho$ the trap frequency in the axial direction $z$. In the experiment with trap frequencies $(\omega_\rho,\omega_z) = 2\pi\times (52.7,1.43)$ Hz the aspect ratio $\lambda$ is small ($\lambda \simeq 0.027$), which causes the cloud of atoms to be cigar-shaped. We condense about $N \simeq 5\times10^7$ atoms and the condensate fraction is $N_0/N \simeq 90\%$. After the preparation the radial trap frequency is modulated with three pulses and the radial breathing mode of the Bose-Einstein condensate is excited. The pulse duration is 50 ms and the modulation depth is 0.125. Due to the superfluid character of the cloud the breathing oscillation is only weakly damped and acts as a drive for higher-order axial excitations. Since the trap frequency in the radial direction greatly exceeds the trap frequency in the axial direction, an excitation in the radial direction can couple through the non-linear interaction to high-order excitations in the axial direction.

\begin{figure}
\includegraphics[scale=1.00]{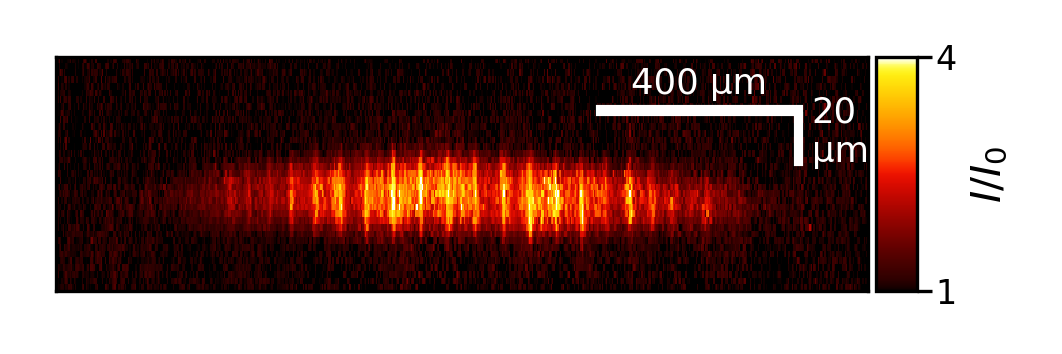}
\caption{Phase contrast image of a higher-order axial excitation. The acquired phase due to the condensate is small with respect to $2 \pi$, such that for a phase delay of $\pi/3$ of the phase spot  the minimum of the intensity is given by $I_0$ and the maximum by $4I_0$, where $I_0$ is the intensity on the camera without the atoms (see Ref.~\cite{pci}). \label{fg:pci}}
\end{figure}

In Fig.~\ref{fg:pci} a typical experimental result is shown 500 ms after the trap modulation with the three pulses using phase contrast imaging~\cite{pci}. Although the intensity in a phase contrast image is not linearly dependent on the density, it is clear that the higher-order excitation as indicated by the fast oscillations in the intensity is mainly in the $z$-direction and that the excitation is almost constant in the radial direction. This is essential for the theoretical model that we construct in this article and allows us to variationally obtain the model from first principles.

\section{Effective action \label{sc:action}}

In the model we describe only the condensed part of the cloud and neglect the influence of the thermal cloud, whose main effect is to introduce a small damping in the condensate dynamics. The state of the condensate is fully described by the wavefunction $\Psi(\vr,t)$ and its evolution is given by the GP equation with a non-linear interaction due to the atom-atom interaction characterized by a single parameter $a_{\rm s}$, the s-wave scattering length.  Equivalently we can use instead an alternative description, where we treat the condensate as a superfluid and use the techniques of quantum hydrodynamics for its description. The condensate is then fully described by the density $n(\vr,t)$ and its phase $\phi(\vr,t)$, for which an effective action can be constructed. Note that both descriptions are identical~\cite{pethick_smith}, but for our purposes it is more appropriate to use the density and the phase. For our bosonic system the effective action is given by
\begin{equation} \label{eq:action}
\cS = \int \dd t \int \dd\vr \; \cL,
\end{equation}
with the Lagrangian density $\cL$ defined by~\cite{Henk}
\begin{eqnarray}
\cL &=& - \hbar n \doubpart{\phi}{t} - n V(\vr)  \label{eq8}  \\ 
& & - \frac{\TtwoB}{2}n^2 -\frac{\hbar^2}{2m}\haak{\frac{(\vn n)^2}{4n}+n(\vn\phi)^2}, \nonumber
\end{eqnarray}
where the quantum pressure is given by the first term within the square brackets. The non-linear interaction is determined by the two-body interaction $\TtwoB=4\pi\hbar^2 a_{\rm s}/m$. It is straightforward to check that minimizing the action with respect to the phase $\phi$, \idest, ${\delta \cS}/{\delta\phi}=0$, corresponds to the continuity equation for the density
\begin{equation}
\doubpart{n}{t}+\vn \cdot \vj=0. \label{eq8a}
\end{equation}
Minimizing the action with respect to the density, \idest, ${\delta \cS}/{\delta n}=0$,  corresponds to the Josephson equation
\begin{equation}
\hbar\doubpart{\phi}{t}+\embr{\frac{1}{2}m \vv^2+V+\TtwoB n-\frac{\hbar^2}{2m\sqrt{n}}\vn^2\sqrt{n}}=0. \label{eq8b}
\end{equation}
Here we introduced the current density $\vj(\vr,t)=n(\vr,t) \vv(\vr,t)$ and the velocity of the flow $\vv(\vr,t)=\hbar\vn \phi(\vr,t)/m$. The characteristic length of the trap is the harmonic oscillator length $a_{\rm ho}$, which is defined as $a_{\rm ho}= \sqrt{\hbar/m\bar{\omega}}$ with the geometric trap frequency $\bar{\omega}=(\omega_x \omega_y \omega_z)^{1/3}$. Note that for the experimental parameters $N_0 a_{\rm s}/a_{\rm ho} \simeq 10.000$ and thus much larger than 1,  such that the quantum pressure term can be neglected, or in the language of the GP equation, that the atoms are in the Thomas-Fermi (TF) limit~\cite{pethick_smith}. 

In our model we assume that the density modulation of the excitations are small and thus that we can expand the total density $n$ and phase $\phi$ as the sum of a condensate part determined by $n_D$ and $\phi_D$ containing the radial dynamics, and the part containing the higher-order axial excitations described by $n_A$ and $\phi_A$:
\begin{equation}
n=n_D+n_A, \qquad \phi=\phi_D+\phi_A.\label{eq9}
\end{equation}
Substituting \eqref{eq9} into the Lagrangian density of \eqref{eq8}, and considering only the second-order and the third-order terms of $n_A$ or $\phi_A$, 
the Lagrangian density becomes
\begin{eqnarray} 
\cL_A  &=&   -\hbar n_A \doubpart{\phi_A}{t} -  \frac{\TtwoB}{2} n_A{}^2   \label{eq11} \\
& - &   \frac{\hbar^2}{2m} \embr{n_D(\vn\phi_A)^2 + 2 n_A \vn \phi_D \cdot \vn \phi_A + n_A (\vn\phi_A)^2 }. \nonumber 
\end{eqnarray}
Here the quantum pressure term is neglected, as discussed after \eqref{eq8b}. Terms depending solely on the properties of the radial excitations $n_D$ and $\phi_D$ have been eliminated,  but they are  reintroduced later on in \scref{back}. 

Since radial excitations have a large energy $\hbar \omega_D \simeq 2\hbar\omega_\rho$ with respect to the splitting of the excitations in the axial direction of about $2 \hbar\omega_z$, we expect that radial excitations can in principle be coupled to many different modes $j$ in the axial direction. Thus we expand the higher-order axial mode density $n_A$ and phase $\phi_A$ in modes $j$ as
\begin{equation}
n_A=\sum_j n_j, \qquad  \phi_A=\sum_j  \phi_j, \label{eq13} 
\end{equation}
where the mode functions associated with the quantum number $j$ are still to be determined.

There has been some theoretical analysis using a time-dependent non-polynomial, non-linear Schr\"odinger equation applied to model cigar-shaped condensates~\cite{Salasnich2002,Nicolin2007}. However, there are exact solutions using Legendre polynomials $P_j(\tz)$ with $\tz=z/R_z$ for an one-dimensional Bose-Einstein condensate in a harmonic trap in the TF regime~\cite{P00,Henk}. Here $R_z$ is the Thomas-Fermi size of the condensate in the axial direction. Under our experimental conditions it is clear that the cloud is not in the one-dimensional regime, since the chemical potential $\mu$ ($\mu/h \simeq$ 2 kHz) is much larger than the excitation energy $\hbar\omega_\rho$.  However, as can be seen from the excitation profile in Fig.~\ref{fg:pci}, there is nearly no radial dependence of the higher-order axial profile, so the Legendre polynomials seem to be the most suitable basis functions for the expansion of the excitation profile.  Furthermore, due to parity conservation the mode number $j$ of $P_j(\tz)$ is required to be even, \idest, the radial breathing mode can only couple to axial modes with an even parity. The higher-order Legendre polynomials have simple sinusoidal behavior around $\tz \simeq 0$, but for $|\tz| \rightarrow 1$ their amplitude strongly increases. In the experiment, the density of the condensate goes to zero near the edge with $|\tz| \simeq 1$, but also the excitation profile diminishes strongly. Thus a single Legendre polynomial does not satisfy the right boundary condition. However, if we subtract two Legendre functions, where the mode numbers differ by two, the linear combination remains a solution for the one-dimensional case and decreases to zero near the edges. Thus we choose the high-order axial mode function $L_j(\tz)$ as  
\begin{eqnarray}\label{eq16}
L_j(\tz)=P_{4j+2}(\tz)-P_{4j}(\tz), \label{eq:Ljz}
\end{eqnarray}
where $j$ is chosen in such a way that the mode functions are countable and $j=0,1,2,3,\ldots$. Note that these mode functions are independent of the time $t$ if the length $R_z$ of the condensate is constant. However, in our experiment, where the radial breathing mode is excited, there is coupling with the axial modes~\cite{Smits2018}, as $R_z$ and thus also $L_j(\tz)$ are functions of time. 

For the density and the phase of the higher-order axial mode functions we can now use the variational {\it Ansatz}
\begin{eqnarray}
n_j(z,t) &\equiv& -\dkap_j(t)L_j(\tz), \label{eq14} \\
\phi_j(z,t) &\equiv& \frac{\TtwoB}{\hbar}\kappa_j(t)L_j(\tz), \label{eq15}
\end{eqnarray}
with $\kappa_j$ the mode amplitude. The relation between the density and phase are taken such that the mode functions obey the Josephson equation of \eqref{eq8b} if the length of the condensate $R_z$ is constant. In \eqref{eq14} and in the remainder of the article the dot indicates the derivative with respect to time. Note that the excitation profile is now fully determined by the time-dependent amplitudes $\kappa_j(t)$. 

For the radial excitation,   the density and the phase are determined in the TF approximation by~\cite{Ka96a,Ka96b,Da97a,Da97b}
\begin{equation}
n_D(\vr,t)=\frac{n_0}{b_x b_y b_z}\embr{ 1-\tx(t)^2-\ty(t)^2-\tz(t)^2} , \label{eq5}
\end{equation}
and
\begin{equation}\label{eq6}
\vn\phi_D(\vr,t)=\frac{m \Ri{\rho}}{\hbar} \embr{\db_x\tx(t)\ve_x + \db_y \ty(t) \ve_y  + \frac{\db_z\tz(t)}{\lambda}\ve_z},
\end{equation}
in which the density in the center of the condensate is given by  $n_0 \equiv m\omega_\rho{}^2 \dingus/2\TtwoB$ with $ \Ri{\rho}\equiv\Ri{x} = \Ri{y}$. The dimensionless time-dependent variables are given by
\begin{eqnarray}
\tx(t) & = & \frac{x}{b_x(t)\Ri{x}},    \qquad \ty(t)=\frac{y}{b_y(t)\Ri{y}}, \label{eq7} \\
\qquad \qquad & & \tz(t)=\frac{z}{b_z(t)\Ri{z}}, \nonumber
\end{eqnarray}
where the parameters $b_i(t)$ with $i=x,y,z$ determine the explicit time-dependence of the size of the condensate and $\Ri{i}$ are the equilibrium values. The evolution of the parameters $b_i$ is discussed in \scref{back}.

To find the evolution of the amplitudes $\kappa_j$ we can substitute \eqsref{eq13}{eq15} in the Lagrangian density of \eqref{eq11} and obtain for the Lagrangian $\nL_A$ defined by $\nL_A=\int \dd \vr\;\cL_A $ the result
\begin{eqnarray}\label{eq20}
\nL_A & = & \eta \sum_{ij} \left( \frac{Q_{ij}}{2 } b_x b_y b_z \embr{ \dkap_i \dkap_j-\Gamma_{ij}(t)\kappa_i \kappa_j} \right. \nonumber\\
& & + \left. \frac{\TtwoB \lambda^2}{2 m \dingus}\frac{b_x b_y}{b_z}\sum_k M_{ijk}\dkap_i \kappa_j \kappa_k \right),
\end{eqnarray}
where we have integrated out the dependence on $\vr$ in \eqref{eq11}, since both the radial and axial modes  are fully specified in the $\rho$ and $z$-direction. The Lagrangian is given in terms of the following integrals
\begin{eqnarray}\label{eq17}
T_{ij} &=& \int_{-1}^{1} \dd \tz (1-\tz^2)^2 L'_i(\tz)L'_j(\tz),\\
Q_{ij} &=& \int_{-1}^{1} \dd \tz (1-\tz^2) L_i(\tz)L_j(\tz),\label{eq18} \\
M_{ijk} &=&\int_{-1}^{1} \dd \tz (1-\tz^2) L_i(\tz)L'_j(\tz)L'_k(\tz), \label{eq19}
\end{eqnarray}
combined with the effective mass parameter $\eta$,  
\begin{equation}\label{eq21}
\eta=\pi \TtwoB \dingus R_z(0),
\end{equation}
and the square of the effective frequency $\Gamma_{ij}(t)$ given by 
\begin{equation}\label{eq22}
\Gamma_{ij}(t)=\frac{\omega_z{}^2}{4}\frac{T_{ij}}{Q_{ij}}\frac{1}{b_x b_y b_z{}^3}.
\end{equation}
In \eqsref{eq17}{eq19} the prime indicates the derivative with respect to $\tz$. Note that the first two terms of the Lagrangian of \eqref{eq20} have the form of a harmonic oscillator, where the first term in the brackets is proportional to the kinetic energy and the second part to the potential energy. From the Lagrangian we can derive the evolution equation of the amplitudes $\kappa_i$ in \scref{diag}, but before proceeding we need to test the validity of the  mode functions $L_i(\tz)$ in our variational {\it Ansatz}.  

\section{Comparison to numerical simulations\label{sc:simul}}

In order to test our mode functions $L_j(\tz)$, we can compare their density profile with the experimental density profiles, as seen for instance in Fig.~\ref{fg:pci}. However, in the experiment we have to excite the axial excitation sufficiently to be able to detect the density modulations, although  in our model we assume that the excitations are small compared to the condensate density. Strong excitations also have the drawback that multiple modes $j$ can be excited simultaneously and that the experimental excitations have to be compared to a linear superposition of modes. This adds a large number of adjustable parameters in the comparison and thus makes the outcome less certain. Finally, as discussed in Ref.~\cite{Smits2018}, due to small imperfections in the magnetic trap, radial excitations couple to the scissor mode, which adds even more uncertainty in the comparison.

As stated in \scref{action} we incorporate only the condensate in our model and its evolution is described by the GP equation. There are very efficient schemes, which allow for the numerical integration of the GP equation using time-splitting methods. In our case with the cylindrical symmetric trap, we can reduce the number of dimension from three to two, which drastically reduces the computation time. Since in the simulation we simulate the (complex) wavefunction, we have not only access to the density, but can also extract the phase of the condensate. This allows for the determination of the flux in the condensate, which on the one hand only depends on the excitations and not on the bulk of the condensate, and on the other hand is a quantity that cannot be obtained experimentally. Therefore in this section we describe the numerical simulations of the GP equation, and the comparison of the outcomes with the theoretical model.

For simulations, the GP equation  is rewritten in cylindrical coordinates $(\rho, z)$ with the condensate wavefunction equal to $\Psi(\vr, t) = f(\rho,z, t) / \sqrt{\rho}$ using the axial symmetry. The equation to solve becomes
\begin{multline}
(i-\hat{\alpha})\hbar \doubpart{f(\rho,z,t)}{t} = \left( -\frac{\hbar^2}{2m} \haak{\doubpart{^2}{\rho{}^2}  + \doubpart{^2}{z^2} -\frac{1}{4 \rho^2} }+\right.\\
\left. V(\rho,z) +\frac{\TtwoB}{2 \pi \rho} \left|f(\rho,z,t)\right|^2 - \mu \right) f(\rho,z,t),
\label{eq:gpe_sim}
\end{multline}
where $\hat{\alpha}$ is a phenomenological damping constant~\cite{Stoof1999}. This equation is solved by employing a splitting spectral method~\cite{bao}. Time steps are chosen to be 0.001$\tau_p$ with $\tau_p=2\pi/\omega_\rho$ the oscillation period in the radial direction. The grid size is 1024 $\times$ 256 with a physical size of $[-2 R_z,2 R_z] \times [-2 R_\rho,2 R_\rho]$. The simulation is initiated with a profile in the TF limit and the ground state is found by imaginary time evolution. After reaching the ground state, the system can be excited in two ways. First, we can employ the excitation scheme of the experiment by modulating the radial trap frequency. This allows for a full comparison with the outcome of the experiments, although in that case the system is excited strongly. Alternatively, we can artificially imprint the proper phase profile corresponding to one mode function $j$ on the condensate, such that the condensate starts to oscillate in the right mode. From these simulations an accurate frequency for each of the eigenmodes can be extracted. The last method is employed in this section.

\begin{figure}[t]
\includegraphics[scale=1.0]{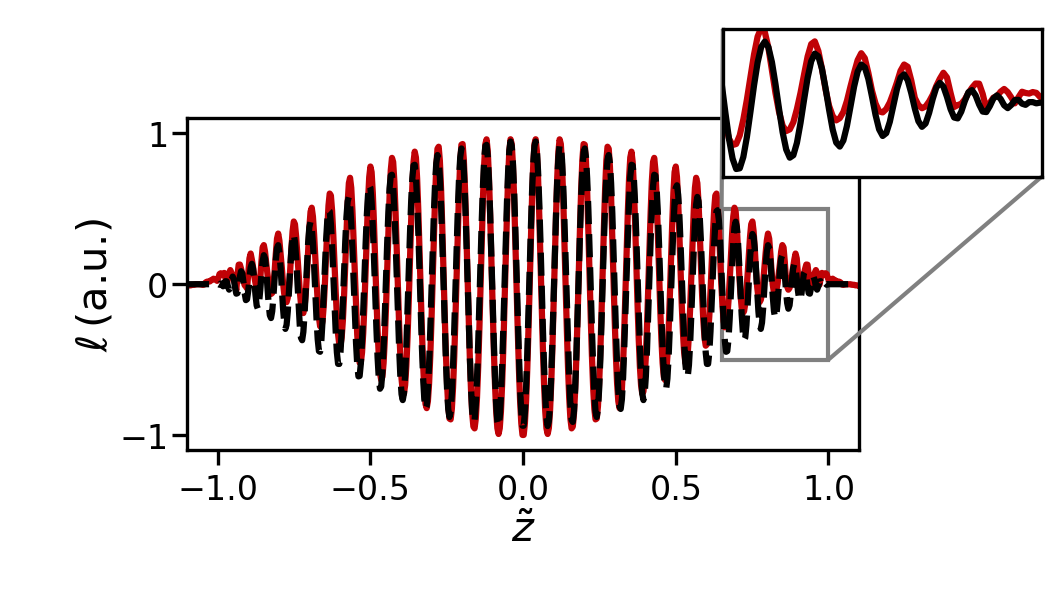}
\caption{(Color online) The high-order line density profile $\ell(\tz)$ as a function of $\tz$ with $j=19$. The black curve is the theoretical prediction of the mode density profile, whereas the red curve corresponds to the numerical simulation. \label{Fig1} }
\end{figure}

In Fig.~\ref{Fig1} we plot the line density profile of the high-order axial mode $\ell(\tz)=\pi (1 - \tz^2) n_A(\tz)$ as a function of $\tz$ by subtracting the ground state of the condensate from the profile after the phase imprinting. The figure shows that our theoretical prediction agrees well with the simulation data and that there are only relative small deviations near the edge of the condensate at $|\tz|\rightarrow1$. Note that due to the quantum pressure term in the simulation the density exponentially decays to zero at the edge, whereas in the model the density has a discontinuity near the edge, which is not physical. This may explain the deviations near the edge of the condensate.

It is also useful to investigate the axial component of the current density $J_z(\tz)$, which is given by 
\begin{equation}\label{eq22a}
J_{z}(\tr,\tz,t)\equiv n(\tr,\tz,t)\frac{\hbar}{m}\doubpart{\phi(\tr,\tz,t)}{z}.
\end{equation}
Using the expansion of \eqref{eq9}, the expressions for  $n_D$, $\phi_D$ in \eqsref{eq5}{eq7}, and $n_A$, $\phi_A$ in \eqsref{eq13}{eq15}, we ultimately obtain
\begin{eqnarray}\label{eq22b}
J_{z}(\tr,\tz,t) &=& \embr{\frac{n_0}{b_x b_y b_z}(1-\tr^2-\tz^2)-\dkap_j L_j(\tz)} \\ \nonumber
&\times& \embr{\Ri{z} \db_z \tz+\frac{\TtwoB \kappa_j}{m \Ri{z} b_z} (8j+3)P_{4j+1}(\tz)}.
\end{eqnarray}
Here the following property of the Legendre polynomials 
\begin{equation}\label{eq22dc}
\doubpart{}{z} \embr{P_{4j+2}({z})-P_{4j}(z)}=(8j+3)P_{4j+1}(z)
\end{equation}
is used. 

\begin{figure}[t]
\includegraphics[scale=1.00]{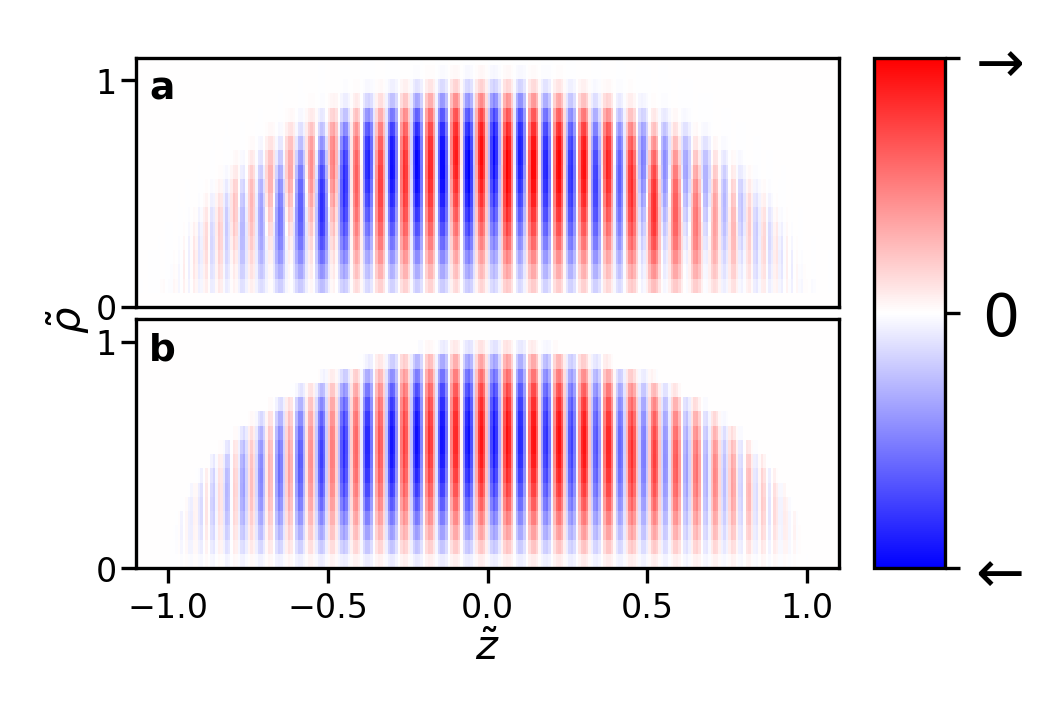}
\caption{(Color online) The density plot of the axial particle flux $\tr J_z(\tr,\tz,t)$ as a function of the scaled radial and axial position for $j=19$ for (\textbf{a}) the numerical simulation and  (\textbf{b}) our theoretical model.\label{Fig2}}
\end{figure}

In Fig. \ref{Fig2} the comparison for the flux between the theoretical model and the simulation is shown. The agreement between theory and simulation is excellent apart from a small deviation near the edge of the condensate. It is also clear that the numerical simulations show only small variations of the flux in the radial direction, proving our basic assumption that the restriction of the mode function to only allow for axial variation is justified. The agreement for both the density and the flux between simulation and model shows that our choice for the mode functions is very accurate. Thus we can now address the dynamics of the high-order axial modes using these mode functions.

\section{No mode coupling between axial modes\label{sc:diag}}

If we exclude the coupling of different higher-order modes, we need to consider only the diagonal terms in the Lagrangian $\nL_A$ in \eqref{eq20} and we obtain 
\begin{eqnarray}\label{eq23}
\nL_{A} &=& \eta \sum_j \left( \frac{Q_{jj}}{2} b_x b_y b_z \haak{\dkap_j{}^2-\Gamma_{jj}(t)\kappa_j{}^2} \right. \nonumber\\
& & + \left. \frac{\TtwoB \lambda^2}{2 m \dingus} \frac{b_x b_y}{b_z} M_{jjj}\kappa_j{}^2\dkap_j \right).
\end{eqnarray}
Because the Lagrangian contains only the first-order time derivative $\dkap_j$, the equations of motions follow from the Euler-Lagrange equation
\begin{eqnarray}\label{eq24}
\doubpart{\nL}{\kappa_j}-\doubfull{}{t} \doubpart{\nL}{\dkap_j}=0.
\end{eqnarray}
Using \eqref{eq24} for the Lagrangian of \eqref{eq23} the equation of motion for the amplitudes of the higher-order axial mode becomes
\begin{eqnarray}\label{eq25}
\ddkap_j &+& \Gamma_{jj}(t)\kappa_j + \embr{\frac{\db_x}{b_x} + \frac{\db_y}{b_y} + \frac{\db_z}{b_z}}\dkap_j \\
& &+\frac{\TtwoB \lambda^2 M_{jjj}}{2 m \dingus Q_{jj}} \embr{\frac{\db_x}{b_x}+\frac{\db_y}{b_y}-\frac{\db_z}{b_z}}\frac{\kappa_j{}^2}{b_z{}^2} =0. \nonumber
\end{eqnarray}
When the radial excitations are small and the terms $\db_i/b_i$ can be neglected in \eqref{eq25}, the solution of \eqref{eq25} is simply sinusoidal with a frequency $\Omega_j$ given by
\begin{equation}\label{eq26}
\Omega_j = \sqrt{\Gamma_{jj}} = \frac{\omega_z}{2} \sqrt{\frac{T_{jj}}{Q_{jj}}} , 
\end{equation}
with $b_x=b_y=b_z=1$. The integrals $T_{jj}$ and $Q_{jj}$ can easily be evaluated and for large $j$ this leads to $\Omega_j \simeq (2j+3/4)\omega_{z}$. So in particular the splitting of the modes is equal to $2 \omega_z$.

\begin{figure}[t]
\includegraphics[scale=1.00]{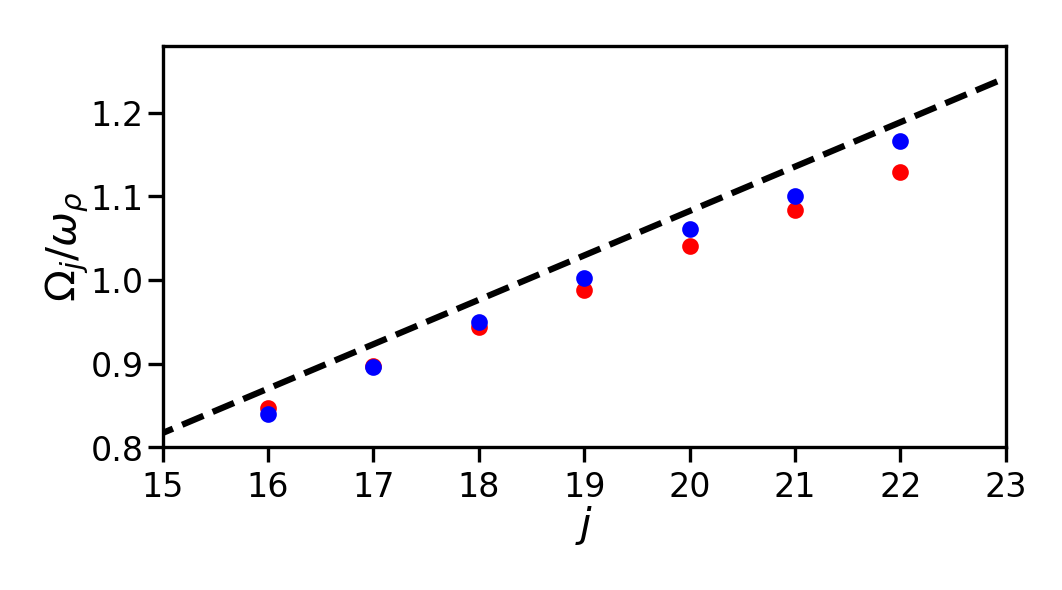}
\caption{(Color online) High-order axial mode frequency $\Omega_j$ in units of $\omega_\rho$ as a function of mode number $j$. The black dashed line is the theoretical prediction of \eqref{eq26}, the red dots are the frequencies for the simulation data by phase-imprinting and the blue dots are the frequencies obtained by the integration of the mode function found from the simulation data (as discussed in the text). \label{Fig3}}
\end{figure}

In Fig.~\ref{Fig3} we show the comparison between the theoretical results of the high-order axial-mode frequency $\Omega_j$ and the numerical simulation through the phase-imprinting method, where the mode frequency is extracted by fitting the oscillatory pattern after the imprinting. The deviations are of the order of 5\% in the interesting region near $j=19$, or alternatively, the mode number $j$ is off by 1. This is acceptable considering the fact that the variational model uses a very small number of parameters to describe the dynamics of the whole superfluid. One possible explanation for the deviations is the axial dependence of the mode profile. To investigate, whether this deviation derives from the difference between the theoretical mode function and mode function in the simulation, we used the mode profile from the simulation to calculate $T_{jj}$ from \eqref{eq17} and $Q_{jj}$ from \eqref{eq18} to calculate an estimate of $\Omega_j$ using \eqref{eq22}. Although these values of $\Omega_j$ are closer to values of our theoretical mode function, they still do not fully agree with the theoretical values, presumably due to our assumption of neglecting the radial dependence in the mode profile.

Finally, the Lagrangian $\nL_A$ of \eqref{eq20} contains non-diagonal elements, which we have neglected until now. In principle, these non-diagonal elements $Q_{ij}$ and $\Gamma_{ij}(t)$ can be included in the analysis.  Applying the Euler-Lagrange equation including these non-diagonal terms leads to coupling of the form $Q_{j\pm 1,j}$ and $\Gamma_{j\pm 1,j}(t)$. By diagonalizing the resulting equations we find new eigenvectors with almost equal eigenvalues, where the nearest neighbor contribution is less than $5\%$ for our experimental parameters. Therefore, we ignore the non-diagonal effects in the remainder of this article.

\section{Back-action\label{sc:back}}

In \scref{action} we have concentrated on the part of the Lagrangian density that involves only the higher-order axial modes. However, there is also dynamics in the condensate in the radial direction, and this can be obtained by considering the full Lagrangian density $\cL$ of \eqref{eq8}, substituting \eqref{eq13} and only retaining terms that are second- or third-order in the density $n_A$ or phase $\phi_A$. We obtain for the Lagrangian
\begin{equation}\label{eq28}
\nL=\nL_D+\nL_A,
\end{equation}
where $\nL_A$ is given by \eqref{eq20} and the Lagrangian $\nL_D$ of the condensate is given by
\begin{eqnarray}\label{eq27}
\nL_D &=& \frac{1}{14} m \dingus N_0 \bigg( \db_x{}^2 + \db_y{}^2 + \frac{1}{\lambda^2}\db_z{}^2 \nonumber\\
& & - \omega_\rho{}^2 \embr{b_x{}^2+b_y{}^2+b_z{}^2} - \frac{2 \dattum}{b_x b_y b_z} \bigg),
\end{eqnarray}
where we applied \eqsref{eq5}{eq7}. Applying the Euler-Lagrange formalism of \eqref{eq24} with respect to the scaling factors $b_i$ for $\nL$ of \eqref{eq28}, the equation of motion for $b_i$ with back action of the higher-order axial modes become
\begin{equation}
\ddot{b}_i + \omega_i{}^2 b_i-\frac{\dittum}{b_x b_y b_z b_i} + \chi \sum_j I_{ij}(t)=0,\label{eq29}
\end{equation}
with $i=\rho,z$ and the back-action parameter $\chi = -7\pi \TtwoB \Ri{z}/(2 m N_0)$ and
\begin{eqnarray} \label{eq32}
I_{\rho j}(t) &=& Q_{jj} b_\rho b_z \dkap_j{}^2+\frac{\TtwoB M_{jjj}}{m \dottum} \frac{b_\rho}{b_z}\kappa_j{}^2 \dkap_j, \label{eq33} \\
I_{z j}(t) &=& \lambda^2 \bigg( Q_{jj} b_\rho{}^2 \dkap_j{}^2-\frac{\TtwoB M_{jjj}}{m \dottum} \frac{b_\rho{}^2}{b_z{}^2}\kappa_j{}^2 \dkap_j \nonumber\\
& & \qquad \qquad  + \frac{\dottum T_{jj}}{2 b_z{}^3} \kappa_j{}^2 \bigg) . \nonumber
\end{eqnarray}
In absence of back action of the axial dynamics, \eqref{eq29} reduces to the ordinary equations for shape oscillations in a condensate that have already been discussed in Refs.~\cite{Ka96a,Ka96b,Da97a,Da97b}. 

\section{The space-time crystal Hamiltonian \label{sc:STX}}

Although theoretical proposals for time crystals have been discussed in the literature for some time now~\cite{Huang,lustig,WenWei,Mizuta,Russo}, the connection with the present experimental schemes in spin systems have not been established firmly, mainly because the underlying physical description is not known. In the experimental scheme using superfluid bosons~\cite{Smits2018}, this mechanism can be described in detail, as we have shown in \scref{action}. One of the open questions in time crystals is their long-term stability. In the spin systems there is a lot of discussion, whether the many-body interactions in the system lead to the stability of the crystal. However, there are also suggestions that time crystals can be stabilized in a so-called prethermal state~\cite{else}. In this section we show that our effective Lagrangian predicts a prethermal state, where the Hamiltonian in a rotating frame becomes time independent. We also identify in this Hamiltonian a symmetry, which is broken in the phase transition to the space-time crystal.

In the Lagrangian of \eqref{eq23} we substitute for the amplitude $\kappa_j$ in the rotating frame of the drive 
\begin{equation}
\kappa_j = \tkap_j e^{-i \omega_D t/2} + \tkap_j{}^* e^{+i \omega_D t/2}, \label{eq:7-kappaj}
\end{equation}
where we have neglected the third-order term proportional to $M_{jjj}$. We assume that the breathing mode is weakly excited, such that $b_z = 1$ and $b_x = b_y = 1 + A_D \cos(\omega_D t)$ and only retain terms up to first order in $A_D$. Since the amplitudes $\tkap_j$ are slowly varying ($|\dtkap_j / \tkap_j| \ll \omega_D$), we neglect terms proportional to $\dtkap_j{}^2$. By substituting \eqref{eq:7-kappaj} in \eqref{eq23} and applying the rotating-wave approximation, we obtain 
\begin{eqnarray}
\nL_A  && =  \eta\!\sum_j\! Q_{jj} \left(  \delta_j \omega_D \tkap_j \tkap_j^* - \frac{A_D \omega_D{}^2}{8} \haak{\tkap_j \tkap_j + \tkap_j{}^* \tkap_j{}^* } \nonumber \right.\\
& &\left.-\frac{i \omega_D}{2} \haak{\tkap_j \dtkap_j{}^* - \tkap_j{}^* \dtkap_j + A_D \embr{\tkap_j \dtkap_j - \tkap_j{}^* \dtkap_j{}^*}} \right), \nonumber \\ \label{eq:7-lagRWA}
\end{eqnarray}
where we used the approximation $ (\omega_D/2)^2 - \Omega_j{}^2 \simeq  \delta_j \omega_D$, with $\delta_j = \omega_D/2 - \Omega_j$. Using the relation between the Hamiltonian and the Lagrangian
\begin{equation}
\cH \equiv \sum_j \embr{\dtkap_j \doubpart{\nL}{\dtkap_j} + \dtkap_j^* \doubpart{\nL}{\dtkap_j{}^*}} - \nL
\label{eq:7-HamilLag}
\end{equation}
the Hamiltonian in the rotating frame becomes
\begin{equation}
\cH = \eta\!\sum_j\! Q_{jj} \embr{ - \delta_j \omega_D \tkap_j \tkap_j{}^* +  \frac{\omega_D{}^2 A_D}{8} 
\haak{ \tkap_j \tkap_j + \tkap_j{}^* \tkap_j{}^* }}. \label{eq:7-HamRWA}
\end{equation}
In the rotating frame the Hamiltonian thus becomes time-independent leading to the stability of the space-time crystal.

In the space-time crystal the time- and space-translation symmetry of the Hamiltonian are broken. To explore the symmetry breaking, it is most convenient to work in second quantization. Note that the condensed state contains phase fluctuations, but the number of condensed atoms is large (of ${\cal O}(\ee{7})$) and the model so far is a mean-field description. However, in this section we quantize the Hamiltonian by replacing the amplitudes $\tkap_j$ and $\tkap_j{}^*$ by the annihilation $\haj$ and creation $\hajd$ operator, respectively. 

For the quantized Hamiltonian the commutation relation between the creation and annihilation operator $\comm{\haj}{\hajd} = 1$ needs to be fulfilled, such that the operator $\hajd \haj$ determines the number of excitations. For the quantization we replace in \eqref{eq:7-HamRWA} $\tkap_j$ and $\tkap_j{}^*$ by 
\begin{equation}
\tkap_j \rightarrow q_j \haj, \qquad \tkap_j{}^* \rightarrow q_j \hajd, \label{eq:7-ajAj}
\end{equation}
where $q_j$ is a $j$-dependent normalization parameter, which is determined from the Lagrangian in \eqref{eq:7-lagRWA} by calculating the canonical momentum $\hpj = {\partial L}/{\partial \dhaj}$ and requiring $\comm{\haj}{\hpj} = i\hbar$. As a result we find
\begin{equation}
q_j = \sqrt{\frac{\hbar}{\eta Q_{jj} \omega_D}},	\label{eq:7-eta}
\end{equation}
which we substitute in the Hamiltonian in \eqref{eq:7-HamRWA} to find the quantized Hamiltonian 
\begin{equation}
\cH = \sum_j \embr{- \hbar \delta_j \hajd \haj + \frac{\hbar\omega_D A_D}{8} \haak{\haj \haj + \hajd \hajd}},
\label{eq:7-HamQuant}
\end{equation}
where the factor in front of the square brackets acts as a driving parameter.

In \eqref{eq:7-HamQuant} the $U(1)$ symmetry $\haj \rightarrow \haj e^{i\vartheta}$ is broken. As a result, when diagonalizing and solving the system using a Bogoliubov transformation, we find that $\qvalue{\haj \haj}$ is non-zero. Physically this describes the squeezing of the probability distributions of the conjugate variables $\hnj$ and $\hfj$ due to the driving process even when $\qvalue{\hnj} = \langle \hfj \rangle = 0$. The Hamiltonian in \eqref{eq:7-HamQuant} does have the $\mathbb{Z}_2$ symmetry $\haj \rightarrow - \haj$. This discrete symmetry is spontaneously broken when the mode $j$ has a large occupation, indicating that a phase transition has occurred to a phase with $\qvalue{\hnj} \neq 0$ and $\langle \hfj \rangle \neq 0$, which we identify as the space-time crystalline phase.  In the rotating frame this phase can exist for arbitrarily long time, which implies that it is a prethermal phase. Note that the symmetry of the Hamiltonian reflects the fact that the (temporal) phase factor of the space-time crystal is determined by the drive only up to a sign. Choosing one particular sign breaks this Ising symmetry and leads to the formation of the space-time crystal. 

\section{Onset of the space-time crystal\label{sc:higher}}

One of the outcomes of the experiment is that the higher-axial modes appear a long time after the excitation~\cite{Alex}. As already discussed is \scref{action}, many modes can be excited, but in the experiment only one or a few modes are observed. In this section we exploit the equation of motion of \eqref{eq25} to find the frequencies, growth rates and threshold for the higher-order axial modes.

In the experiment, there are certain dissipative processes involved, which have not been included in the model so far. One of the most important dissipation mechanisms occurs through the thermal cloud, although its density is small compared to the condensate. As shown in Ref.~\cite{damp}, by moving through the thermal cloud, excitation in the condensate can be induced, which lead to damping of its motion. We include the damping in \eqref{eq25} by adding a phenomenological damping parameter $\alpha$,
\begin{equation}\label{eq50b}
\ddkap_j + \Gamma_{jj}(t) \kappa_j + \embr{\frac{\dot{b_x}}{b_x}+\frac{\dot{b_y}}{b_y}+\frac{\dot{b_z}}{b_z}+\alpha} \dkap_j =0,
\end{equation}
which assumes that the damping is proportional to the velocity of the condensate. Here we have again neglected the third-order term proportional to $M_{jjj}$ in \eqref{eq25} for simplicity.

\begin{figure}[t]
\includegraphics[scale=0.9]{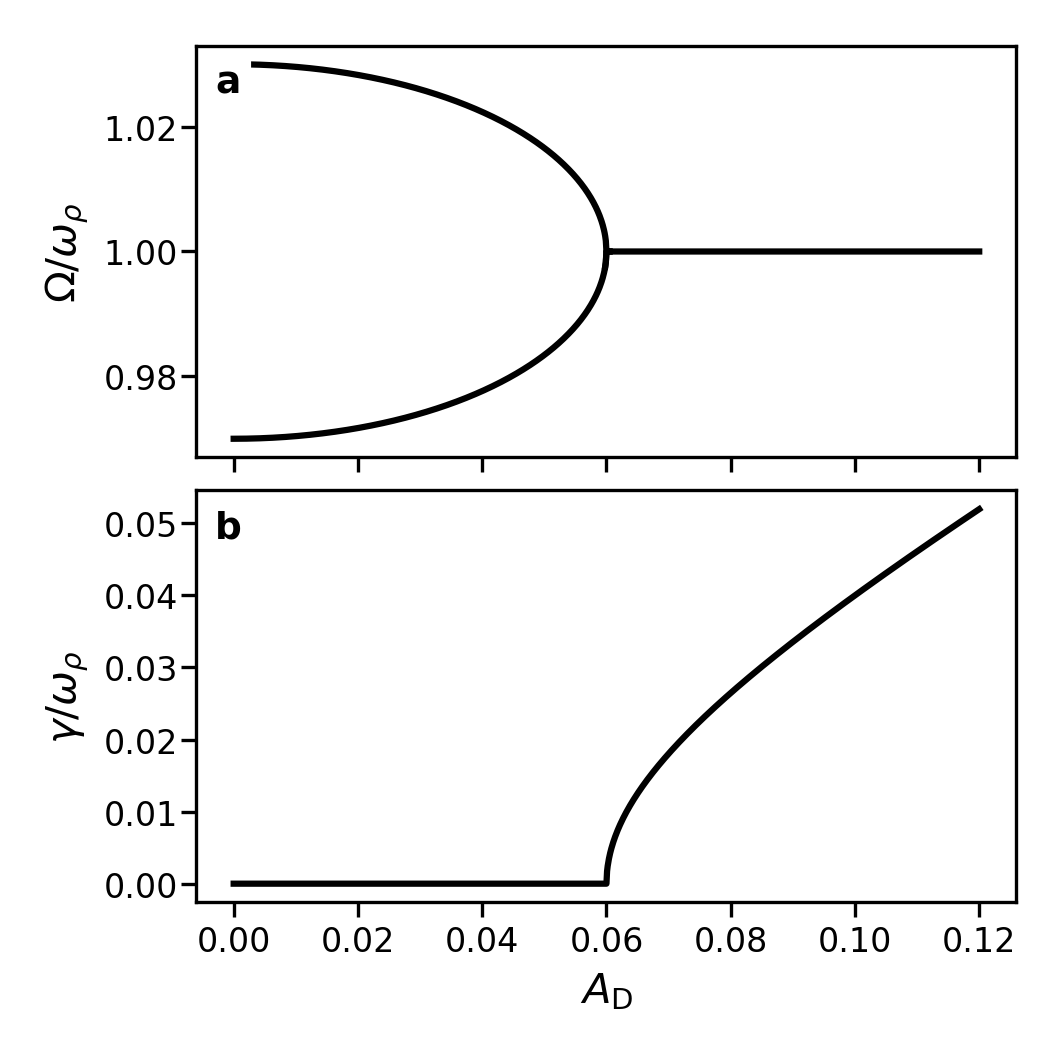}
\caption{(Color online)  \textbf{(a)} The frequency $\Omega=\omega_D/2+\Delta\Omega_j$ and \textbf{(b)} the growth rate  $\gamma$ in units of $\omega_\rho$  for $j$ = 19 as a function of $A_D$ for $\Omega_j=1.03\omega_\rho$ and $\omega_D\simeq 2 \omega_\rho$. There is no damping ($\alpha$ = 0). \label{Fig5}}
\end{figure}

Since the higher-order axial mode is driven at the radial breathing mode frequency $\omega_D$, and the higher-order axial mode oscillates at nearly one half of this frequency, we substitute \eqref{eq:7-kappaj} in \eqref{eq50b} and only retain non-oscillating terms, which corresponds to the rotating-wave approximation of \scref{STX}. Since the mode amplitude is nearly constant during one oscillation of the drive, the second-order derivative of $\ddtkap_j(t)$ can safely be ignored and we obtain
\begin{equation}\label{eq51}
\embr{-\delta_j -i \frac{\alpha}{2} } \tkap_j +A_D \frac{\omega_D}{4} \tkap_j{}^*  =i \dtkap_j.
\end{equation}
Here we have neglected the back action of the higher-order axial modes on the breathing mode as discussed in \scref{back} and thus assumed $A_D$ to  be constant. The solutions are given by $\tkap_j \propto \exp(-i \Delta \Omega_j t)$ with
\begin{equation} \label{eq:freq}
\Delta \Omega_j =  - i \alpha/2 \pm \sqrt{\delta_j{}^2 -(A_D \omega_D/4)^2}  . 
\end{equation}
In case that the expression under the square-root is positive, the solutions (apart from the term proportional to $\alpha$) are oscillatory and the frequencies are shifted up and down with respect to $\omega_D/2$. However, if this expression is negative, the frequencies becomes complex and thus lead to gain of the amplitude. There is only gain, when the detuning is not too large with respect to $\omega_D$:
\begin{equation}
|\delta_j| < \frac{\omega_D A_D}{4}  .
\end{equation}
Here we assumed that $\alpha$ is much smaller than $\delta_j$. In general, there is gain if the growth due to the driving exceeds the damping and the driving amplitude must be larger than the threshold value $A_D^{\rm Th}$, which is given by
\begin{equation}
A_D^{\rm Th} =\frac{ 4 \sqrt{\delta^2 + (\alpha/2)^2}}{\omega_D}.
\end{equation}
In Fig.~\ref{Fig5} we show the results for $\alpha=0$ and non-zero detuning $\delta$. Below the threshold there are two mode frequencies and their splitting decreases if the driving amplitude increases. Above threshold, where the two frequencies coincide, the mode amplitude increases exponentially and the gain coefficient increases nearly linearly with driving amplitude $A_D$. Note that above threshold the oscillation frequency becomes $\omega_D/2$, which implies that the periodicity of the space-time crystal is twice the periodicity of the drive and the discrete time symmetry is broken.

\begin{figure}[t]
\includegraphics[scale=1.0]{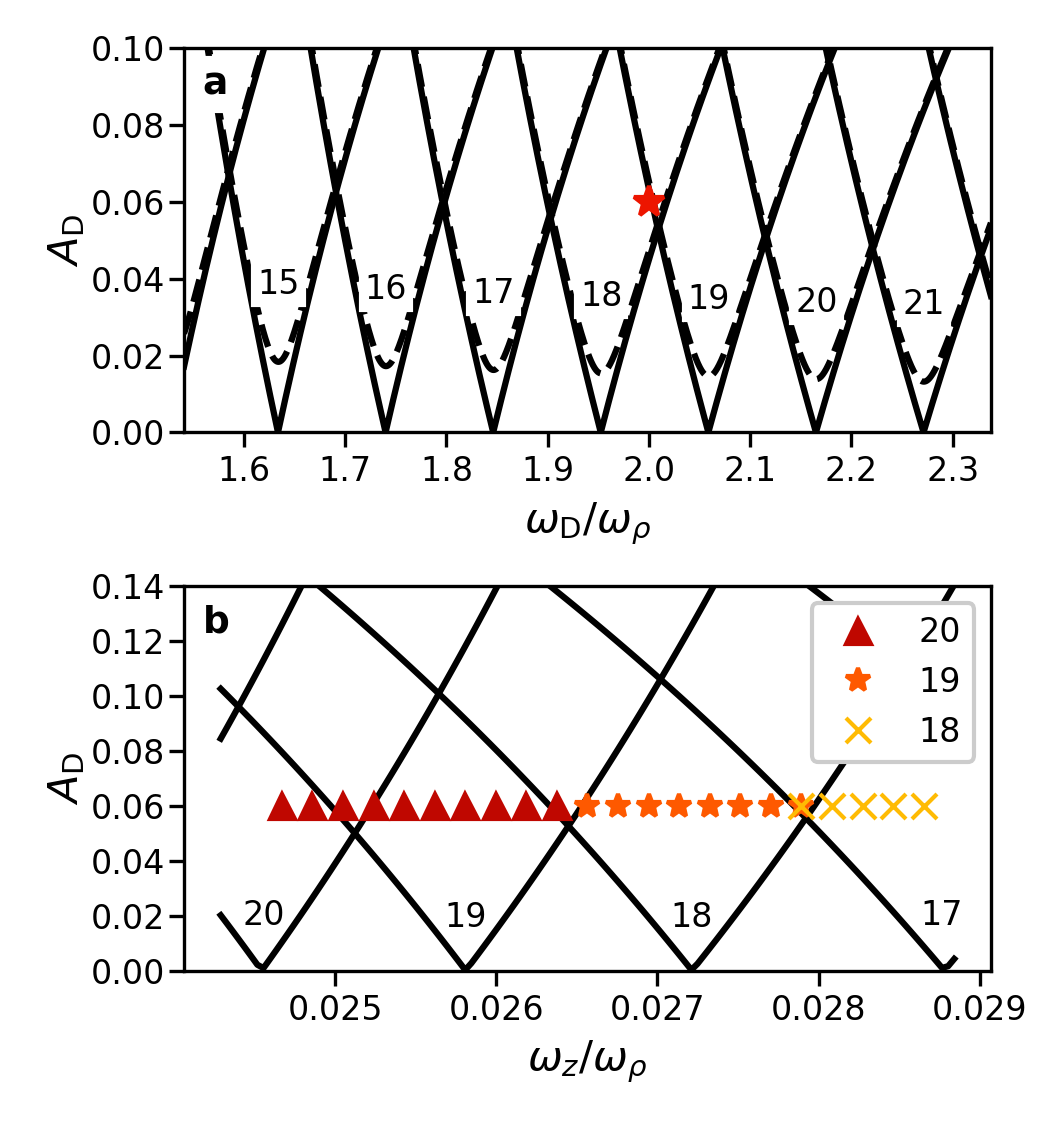}
\caption{(Color online). Linear-response analysis of (\textbf{a}) the threshold of the breathing mode amplitude $A_D^{\rm Th}$ as a function of the radial mode frequency $\omega_D$ and (\textbf{b}) as a function of the axial trap frequency $\omega_{z}$. The results are without damping (solid line) and with damping (dashed line). The red star in (\textbf{a}) indicates the breathing mode amplitude of the simulations of \scref{simul}. The symbols in (\textbf{b}) indicate the dominant high-order axial mode of the fit for the simulation.\label{Fig4}}
\end{figure}

In Fig.~\ref{Fig4}a we have plotted the threshold as a function of the driving frequency $\omega_D$. In the absence of damping the threshold is zero at resonance and the higher-order axial mode can always be excited. However, for non-zero damping, which coincides with any realistic experiment,  there is a minimum requirement of the driving amplitude, which depends also on the detuning of the mode frequency $\Omega_j$ with half the driving frequency $\omega_D/2$. Note that the detuning is not easily adjustable in the experiment, since it depends on $\Omega_j$, which through \eqref{eq26} depends in a complicated way on various parameters of the experiment. Increasing the driving amplitude too much causes heating of the condensate and thus leads to a strong reduction of the number of condensed atoms. This makes the observation of the space-time crystal in the laboratory non-trivial, since there is a small window for the driving amplitude for the observation of the higher-order mode. In Fig.~\ref{Fig4}a we have indicated the conditions for the simulations of \scref{simul} and from Fig.~\ref{Fig4}a it can be seen that the theoretical model predicts the excitation of predominantly the mode with $j$ = 18. 

In Fig.~\ref{Fig4}b we have plotted the threshold as a function of the axial trap frequency, where the radial trap frequency and thus the driving frequency is kept constant. We have run the simulations using phase imprinting under the same trapping conditions and analyzed the axial flux of the higher-order axial mode using different mode numbers $j$. We find the dominant mode from a fit to the simulations by determining the mode $j$, for which the amplitude in the fit is the largest and simultaneously the reduced chi-squared of the fit is minimal. As can be seen from the fit, the dominant mode is close to the theoretical prediction, although the quantum number $j$ seems to off by 1. This presumably has the same cause as the shift in $j$ in the frequency of the mode (see \scref{simul}).

On resonance ($\Omega_j = \omega_D/2$), the detuning $\delta_j=0$ and the growth rate can easily be determined from \eqref{eq:freq}:
\begin{equation}
\gamma = \frac{A_D \omega_D}{4} - \frac{\alpha}{2}.
\end{equation}
Hence, for sufficiently small damping the growth rate becomes $\gamma \simeq A_D\omega_D/4$. From the simulations we have determined the growth rate and we typically find values, which are a factor 3 larger. However, in the simulation the excitation is rather strong and presumably no longer in the linear regime, where our model applies.

\section{Conclusion and outlook}

In conclusion, we have constructed a variational model that describes  the coupling between the radial breathing mode and the higher-order axial modes in a superfluid Bose gas. We have compared the mode profile and the flux of the higher-order modes with numerical simulations and found good agreement.  The coupling between these modes leads to the space-time crystal that has recently been observed. Since our model has been constructed from first principles, it allows to investigate the necessary requirements to observe the space-time crystal. In particular, we have identified an Ising-type symmetry breaking, where the symmetry in the sign of the higher-order mode amplitude is broken. The model is used to predict the onset and growth rate of the space-time crystal and we will compare the outcome of the model with the results of experiments in the near future~\cite{Alex}.

\strut\\[5mm]

\section*{Acknowledgments}
This work is supported the Stichting voor Fundamenteel Onderzoek der Materie (FOM) and is part of the D-ITP consortium, a program of the Netherlands Organization for Scientific Research (NWO) that is funded by the Dutch Ministry of Education, Culture and Science (OCW). The work of L.L. is supported  by the China Scholarship Council (CSC).


\end{document}